\documentclass[a4paper,11pt]{article}
\usepackage{pos}
\usepackage{booktabs}
\usepackage{multirow}
\usepackage[english]{babel}
\usepackage{color}
\usepackage{hyperref}
\usepackage{dcolumn}
\usepackage{bm}
\usepackage{mathtools,braket,blkarray}
\usepackage{subfig}
\usepackage[sort&compress]{natbib}
\newcommand{\MYhref}[3][redLinks]{\href{#2}{\color{#1}{#3}}}%
\definecolor{greenLinks}{rgb}{0, 0.6, 0} 
\definecolor{blueLinks}{rgb}{0, 0, 0.6}
\definecolor{redLinks}{rgb}{0.6, 0, 0}
\definecolor{tempText}{rgb}{0.55, 0.10,0.67}
\definecolor{eprintLinks}{rgb}{0.4, 0.4, 0.4}
\definecolor{journalLinks}{rgb}{0.255, 0.639, 0.647}
\definecolor{TitleBlue}{rgb}{0.255, 0.639, 0.647}
\def\vev#1{\left\langle #1\right\rangle}
\title{Neutrino mass hierarchy from the discrete dark matter model}

\author[a]{Cesar Bonilla}
\author[b]{Johannes Herms}
\author*[c]{Omar Medina}
\author[d]{Eduardo Peinado}

\affiliation[a]{Departamento de Física, Universidad Católica del Norte,\\
Avenida Angamos 0610, Casilla 1280, Antofagasta, Chile}
\affiliation[b]{Max-Planck-Institut f\"{u}r Kernphysik,\\ Saupfercheckweg 1, 69117 Heidelberg, Germany}
\affiliation[c]{Institut de F\'{i}sica Corpuscular
  (CSIC-Universitat de Val\`{e}ncia), AHEP Group,\\ Parc Cient\'ific de Paterna,
 C/ Catedr\'atico Jos\'e Beltr\'an, 2 E-46980 Paterna (Valencia), Spain}
 \affiliation[c]{Instituto de F\'{\i}sica, Universidad Nacional Aut\'onoma de M\'exico, \\ A.P. 20-364, Ciudad de M\'exico 01000, M\'exico.}

\emailAdd{Omar.Medina@ific.uv.es}

\abstract{We explore a possible explanation for the hierarchy in scale between the atmospheric and solar neutrino mass differences ($\lvert  \Delta m^{2}_{31} \rvert$, and $\Delta m^{2}_{21}$)
through the presence of two distinct neutrino mass mechanisms from tree- and one-loop-level contributions. We demonstrate
that the ingredients needed to explain this hierarchy are present in the minimal discrete dark matter model \cite{Bonilla:2023pna}. This scenario is characterized by adding new RH neutrinos and scalar $SU(2)$ doublets to the Standard Model as triplet representations of an $A_4$ flavour symmetry. The $A_4$ symmetry breaking, which occurs at the electroweak scale, leads to a residual $\mathbb{Z}_2$ symmetry responsible for the dark matter stability and dictates the neutrino phenomenology. We show that CP breaking in the scalar potential is needed to fit the neutrino mixing angles. }

\FullConference{%
  8th Symposium on Prospects in the Physics of Discrete Symmetries (DISCRETE 2022)\\
  7-11 November, 2022\\
  Baden-Baden, Germany
}

\begin{document}
\maketitle

\section{Introduction}
\label{sec:Introduction}
The origin of neutrino mass and the unknown nature of dark matter (DM) are strong evidence of physics beyond the standard model (BSM). Moreover, unscrambling the flavour puzzle present in the SM is essential for building a fundamental theory of particle interactions. It is so because most of the input parameters in the SM are directly related to flavour. With the unprecedented reduction of experimental uncertainties in the measurement of lepton mixing parameters \cite{deSalas:2020pgw}, it has become increasingly compelling to explore models where these three issues are addressed together.

\begin{figure}
\centering
\includegraphics[width=5.5in]{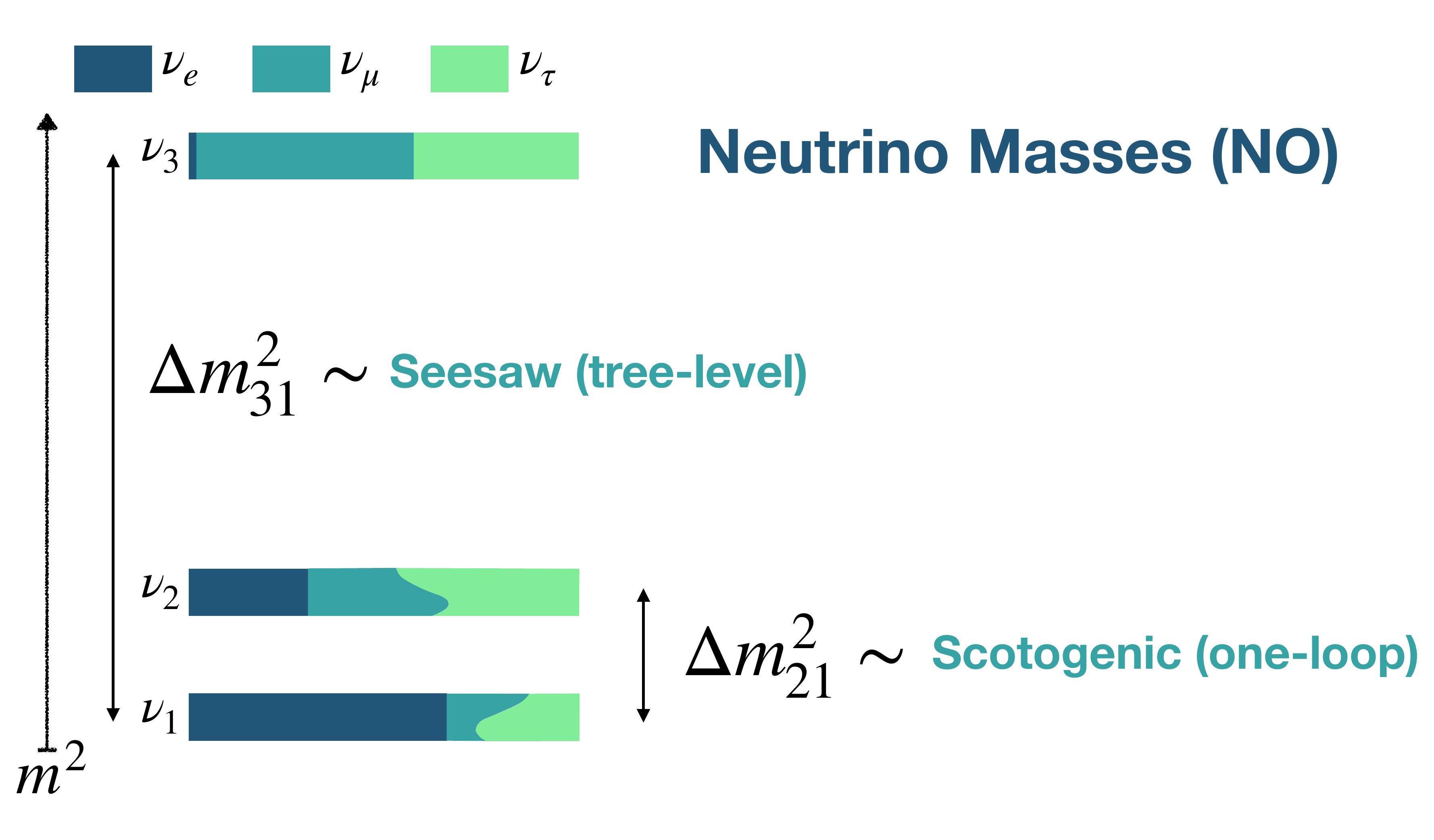}
    \caption{
   In the discrete dark matter (DDM) model \cite{Bonilla:2023pna}, the neutrino mass splittings $\lvert  \Delta m^{2}_{31} \rvert$, and $\Delta m^{2}_{21}$ are generated; by a seesaw, and a scotogenic mechanism respectively \cite{Schechter:1980gr,Ma:2006km,Escribano:2020iqq}. This figure illustrates how this model can naturally explain the hierarchy between these mass splittings and the neutrino mass ordering through the dominance of a tree-level mass mechanism over a loop-suppressed one.}
    \label{fig:ScotoSeesaw}
\end{figure}

In this work, based on \cite{Bonilla:2023pna}, we summarize the interesting features of the discrete dark matter model (DDM)  \cite{Hirsch:2010ru,Boucenna:2011tj}, which deals with the three issues mentioned above. This model is a SM extension with an $A_4$ flavour symmetry whose spontaneous breaking stabilizes a DM scalar particle candidate. We demonstrate that in this model, an interplay between two mass mechanisms (tree-level seesaw and one-loop scotogenic) emerges naturally from the $A_4$ symmetry and its breakdown. This feature of the model predicts Normal Ordering (NO) of neutrino masses and explains the hierarchy between the atmospheric $\lvert \Delta m^{2}_{31} \rvert\sim 2.55\times 10^{-3}$ (NO), and the solar  $\Delta m^{2}_{21}\sim 7.5\times 10^{-5}$ mass splittings, which differ by two orders of magnitude~\cite{deSalas:2020pgw}. This is illustrated in Figure \ref{fig:ScotoSeesaw}.

This work is structured as follows. In Section \ref{sec:DDM} we outline the fields and symmetries of the model.  In Section \ref{sec:NeutrinoMasses} we describe the neutrino mass mechanisms in the model. In Section \ref{sec:NormalOrdering} we summarize our results, and lastly Section \ref{sec:Conclusions} contains our conclusions.
\section{The discrete dark matter model}
\label{sec:DDM}
This model extends the SM symmetry group by an $A_4$ flavor symmetry; and the field content by three  scalar $SU(2)_L$ doublets $\eta = (\eta_1,\, \eta_2,\, \eta_3)$, and three right-handed (RH) neutrinos $N= (N_1, N_2, N_3)$; both as $A_4$ triplets \cite{Bonilla:2023pna,Hirsch:2010ru,Boucenna:2011tj}.
The quantum numbers for the fields in the model are summarized in Table~\ref{tab:Mod}.
\begin{table}[h!]
\begin{center}
\begin{tabular}{@{}cccccccccc@{}}
\toprule
            & $L_{e}$                      & $L_{\mu}$      & $L_{\tau}$            & $l_e$                        & $l_{\mu}$    & $l_{\tau}$            & $N$          & $H$           & $\eta$        \\ \midrule
$SU(2)_{L}$ & $\mathbf{2}$                 & $\mathbf{2}$   & $\mathbf{2}$          & $\mathbf{1}$                 & $\mathbf{1}$ & $\mathbf{1}$          & $\mathbf{1}$ & $\mathbf{2}$  & $\mathbf{2}$  \\
$U(1)_Y$    & $-\frac{1}{2}$               & $-\frac{1}{2}$ & $-\frac{1}{2}$        & $-1$                         & $-1$         & $-1$                  & $0$          & $\frac{1}{2}$ & $\frac{1}{2}$ \\
$A_4$       & $\mathbf{1}^{\prime \prime}$ & $\mathbf{1}$   & $\mathbf{1}^{\prime}$ & $\mathbf{1}^{\prime \prime}$ & $\mathbf{1}$ & $\mathbf{1}^{\prime}$ & $\mathbf{3}$ & $\mathbf{1}$  & $\mathbf{3}$  \\ \bottomrule
\end{tabular}
\caption{\it Relevant fields and quantum numbers of the model.} 
\label{tab:Mod}
\end{center}
\end{table}
\subsection{The Yukawa Sector}
The lepton sector Yukawa Lagrangian of the model is 
\begin{equation}
\mathcal{L}^H_{\text{Yukawa}} =
y_e \overline{L}_e l_e H+y_{\mu} \overline{L}_{\mu} l_{\mu} H +y_{\tau} \overline{L}_{\tau} l_{\tau} H + \textit{h.c.},
\label{eq:YukH}
\end{equation}
\begin{equation}
\mathcal{L}^{\eta}_{\text{Yukawa}} =
 y_1^\nu \overline{L}_e (N\, \tilde{\eta})_{\bf{1}^{\prime \prime}} + y_2^\nu \overline{L}_\mu (N\,  \tilde{\eta}) _{\bf{1}} + y_3^\nu \overline{L}_\tau (N\,  \tilde{\eta})_{\bf{1}^{\prime}} + \frac{1}{2}m_{N}\overline{N^c} N+ \textit{h.c.},
 \label{eq:Yuketa}
\end{equation}
where $ \tilde{\eta}=i \tau_2 \eta^\dagger$, and the subscript of the parenthesis denotes the $A_4$ contraction of two triplets, see \cite{Bonilla:2023pna}  for details about the $A_4$ basis considered. Note from the last equation that due to the flavor symmetry, the three RH-neutrinos are degenerate, with mass 
 $m_N$.
 
 Since the charged leptons only couple to $H$, their mass matrix is diagonal and proportional to its vacuum expectation value (VEV), which we label by $v_H$.
\begin{equation}
\label{eq:MDirA}
\frac{v_H}{\sqrt{2}} Y^H_{l} = 
 \frac{v_H}{\sqrt{2}}   \begin{pmatrix}
  y_e  & 0 & 0 \\
 0 & y_{\mu}  & 0  \\
  0 & 0  & y_{\tau} 
  \end{pmatrix}.
\end{equation}
From Eq. (\ref{eq:Yuketa}) the Yukawa coupling matrices of the $\eta$ fields with neutrinos are
\begin{equation}
  Y^{\eta_1}=\begin{pmatrix}
  y_1^\nu  & 0 & 0 \\
  y_2^\nu  & 0 & 0 \\
  y_3^\nu  & 0 & 0 
  \end{pmatrix}, \quad
   Y^{\eta_2}=\begin{pmatrix}
  0&y_1^\nu \omega^2 & 0 \\
  0&y_2^\nu  & 0 \\
  0&y_3^\nu \omega  & 0 
  \end{pmatrix}, \quad
  Y^{\eta_3}=\begin{pmatrix}
  0&0&y_1^\nu \omega   \\
  0&0&y_2^\nu  \\
  0&0&y_3^\nu \omega^2 
  \end{pmatrix}, \quad \text{with}\quad  \omega=e^{i\frac{2\pi}{3}}.
\label{eq:YukEtaCouplings}
\end{equation}
We want to remark that there are only three parameters $y^{\nu}_1$, $y^{\nu}_2$, and $y^{\nu}_3$, ruling the Yukawa interaction of neutrinos. 
As stated in the next section (for a detailed account, see \cite{Bonilla:2023pna}), neutrino masses are generated by a type-I seesaw and a scotogenic mechanism  \cite{Schechter:1980gr,Ma:2006km,Escribano:2020iqq}. The key feature of this model, in contrast to previous works, is that it naturally predicts the dominance of the seesaw over the scotogenic mechanism \cite{Ibarra:2011gn,Rojas:2018wym,Aranda:2018lif,Barreiros:2020gxu,Barreiros:2022aqu,Mandal:2021yph,Ganguly:2022qxj}.

\subsection{Flavor Symmetry Breakdown}
The $A_4$ symmetry is spontaneously broken at the electroweak scale by the $\eta$ field. Furthermore, in the case that only $\eta_1$ acquires a VEV; there is a residual $\mathbb{Z}_2$ symmetry that stabilizes a dark matter particle candidate~\cite{Hirsch:2010ru}. The VEV alignment of this model
\begin{equation}
\vev{\eta^0}=\frac{1}{\sqrt{2}}\begin{pmatrix} v_{\eta_1} \\ 0 \\0 \end{pmatrix}, \qquad \vev{H^0}=\frac{v_H}{\sqrt{2}}, \qquad \text{with} \quad v^2 = v^2_{\eta_1} + v^2_{H} \approx (246 \hspace{2pt} \text{GeV}),
\label{eq:VEVs}
\end{equation}
remains invariant under one of the generators of $A_4$, which generates the remnant symmetry (the details are given in \cite{Bonilla:2023pna}).
The denominated ``dark'' fields, odd under this $\mathbb{Z}_2$, are
\begin{equation}
\label{eq:DarkFields}
\mathbb{Z}_2: \quad \eta_2 \longrightarrow -\eta_2, \quad \hspace{6pt}\eta_3 \longrightarrow -\eta_3, \quad N_2 \longrightarrow -N_2, \quad N_3 \longrightarrow -N_3,
\end{equation}
while the rest of the fields are invariant, including the SM particles, and are referred to as ``active'' fields.
\section{Neutrino mass mechanisms}
\label{sec:NeutrinoMasses}
\begin{figure}[htb]
\centering
\includegraphics[width=3.0in]{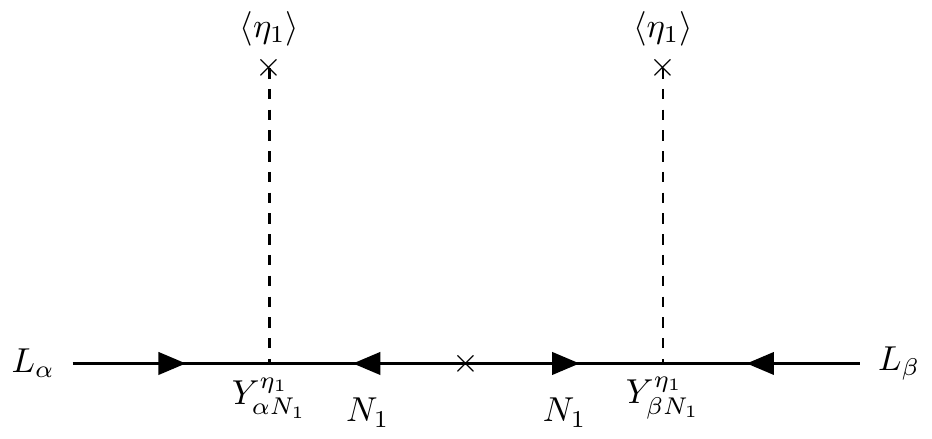}
    \caption{
    The active fields $N_1$ and $\eta_1$ take part in the seesaw mechanism. Their contribution can only generate a rank-1 mass matrix for light neutrinos. Together with its one-loop corrections in Figure \ref{fig:Active-One-Loop} they generate the atmospheric mass splitting $\lvert \Delta m^{2}_{31} \rvert$ (NO).}
    \label{fig:ActiveSeesaw}
\end{figure}

\begin{figure}[htb]%
    \centering
    \subfloat[\centering One-loop correction from the active fields to the seesaw mechanism]{{ 
        \includegraphics[width=2.6in]{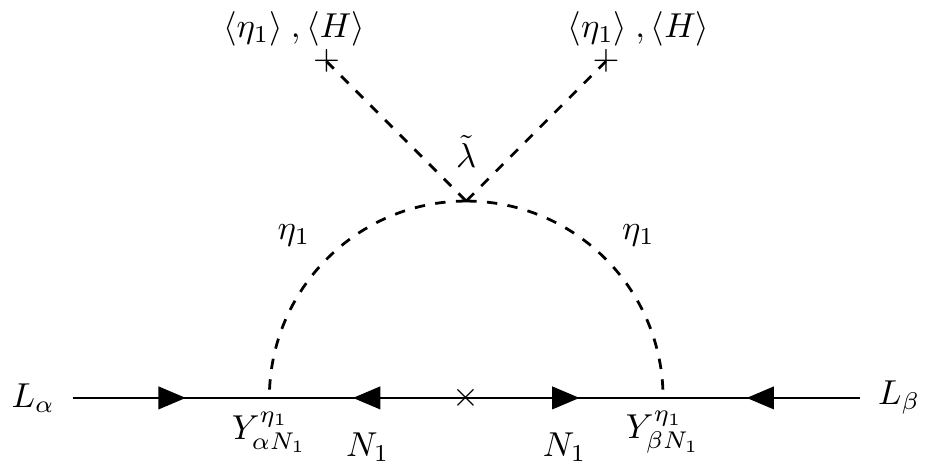}
}}%
    \qquad
    \subfloat[\centering 
    One-loop correction from the Z-boson to the seesaw mechanism]{{
     \includegraphics[width=2.8in]{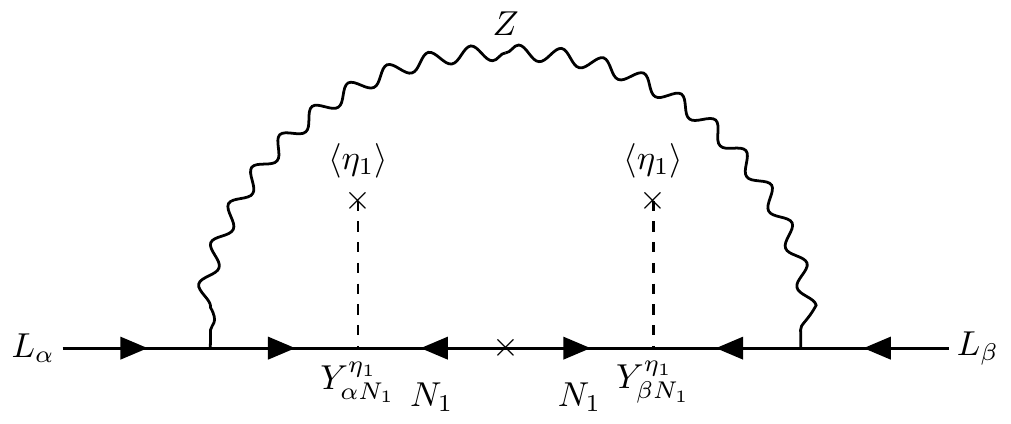}
}}%
    \caption{
    Corrections from the active fields $\eta_1$, $N_1$, and the $Z$-boson to the seesaw mechanism. The $\tilde{\lambda}$ represents the relevant scalar potential couplings. The scalar potential is written explicitly in \cite{Bonilla:2023pna}.}%
    \label{fig:Active-One-Loop}%
\end{figure}

In this model, both the active and dark sectors contribute to generate the neutrino mass matrix. For the sake of clarity we will write their contributions, up to one-loop, separately:
\begin{equation}
(m_{\nu})_{\alpha \beta}=(m^{\text{Active}}_{\nu})_{\alpha \beta}+(m^{\text{Dark}}_{\nu})_{\alpha \beta}, \qquad \text{with} \quad \alpha=e,\mu,\tau.
\label{eq:TotalNeutrinoMassMatrix}
\end{equation}
The fields $\eta_1,~ H,~ N_1$ and  $Z$ contribute to $m^{\text{Active}}_{\nu}$ through a type-I seesaw and its one-loop corrections, as it is illustrated by the Feynman diagrams in Figures \ref{fig:ActiveSeesaw}, and \ref{fig:Active-One-Loop}. Furthermore their total contribution is only a rank-1 matrix, generating thus only the heaviest neutrino mass, as depicted in Figure  \ref{fig:ScotoSeesaw} .

\begin{figure}
    \centering    \includegraphics[width=3.5in]{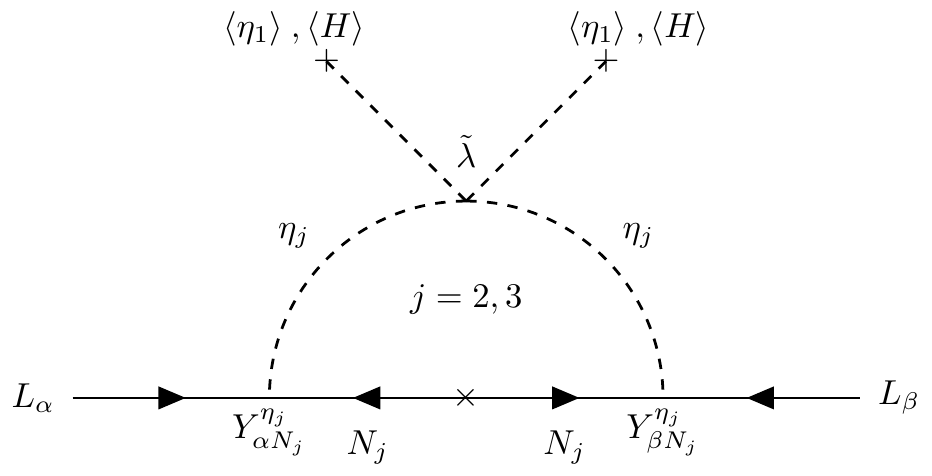}
    \caption{
     The dark fields $N_2$, $N_3$, $\eta_2$, and $\eta_3$ partake in the scotogenic mechanisms, the Yukawa couplings $Y^{\eta_j}_{\alpha N_j}$ for $j= 2,3$ are written explicitly in Eq. (\ref{eq:YukEtaCouplings}). The contribution from this diagram generates a rank-2 mass matrix for light neutrinos. This contribution generates radiatively the solar mass splitting $\Delta m^{2}_{21}$ (NO).
    }
    \label{fig:DarkScotogenic}
\end{figure}

The dark fields fields in Eq. (\ref{eq:DarkFields}) contribute to $m^{\text{Dark}}_{\nu}$ through a scotogenic mechanism, as it is illustrated by the Feynman diagram in Figures \ref{fig:DarkScotogenic}. This mechanism yields a rank-2 matrix, generating the two lighter neutrino masses, also as depicted in Figure  \ref{fig:ScotoSeesaw}. The expressions for these contributions are given in \cite{Bonilla:2023pna}.


\section{Normal Ordering of Neutrino Masses}
\label{sec:NormalOrdering}
Due to the $A_4$ symmetry both mass mechanism in the model share the same neutrino Yukawa couplings. As a consequence, this model predicts normal ordering (NO) for neutrino masses since the scotogenic  rank-2 mass matrix $m^{\text{Dark}}_{\nu}$, cannot dominate over $m^{\text{Active}}_{\nu}$, which emerges at tree-level. Hence, results for inverted order (IO) are not attainable.
\begin{equation}
m^{\text{Active}}_{\nu} \quad \Longrightarrow \quad \Delta m^{2}_{31} \quad \text{(NO)}, \qquad m^{\text{Dark}}_{\nu} \quad \Longrightarrow \quad \Delta m^{2}_{21} \quad \text{(NO)}.
\label{eq:NuScales}
\end{equation}
 This comes as a result of the fact that the mass mediators for both mechanisms in this model belong to the same multiplets of the $A_4$ flavour symmetry.

\subsection{Lepton CP phase $\delta^{\ell}$, and the lightest neutrino mass }
\begin{figure}[htb]%
    \centering
    \subfloat[ The regions in different shades of green  are the $90$, $95$, and $99\%$ CL regions in the $\sin^{2}\theta_{23}$ vs. $\delta^{\ell}$ plane \cite{deSalas:2020pgw}. Table \ref{tab:fit} displays the benchmark point data.]{{ 
        \includegraphics[width=2.75in]{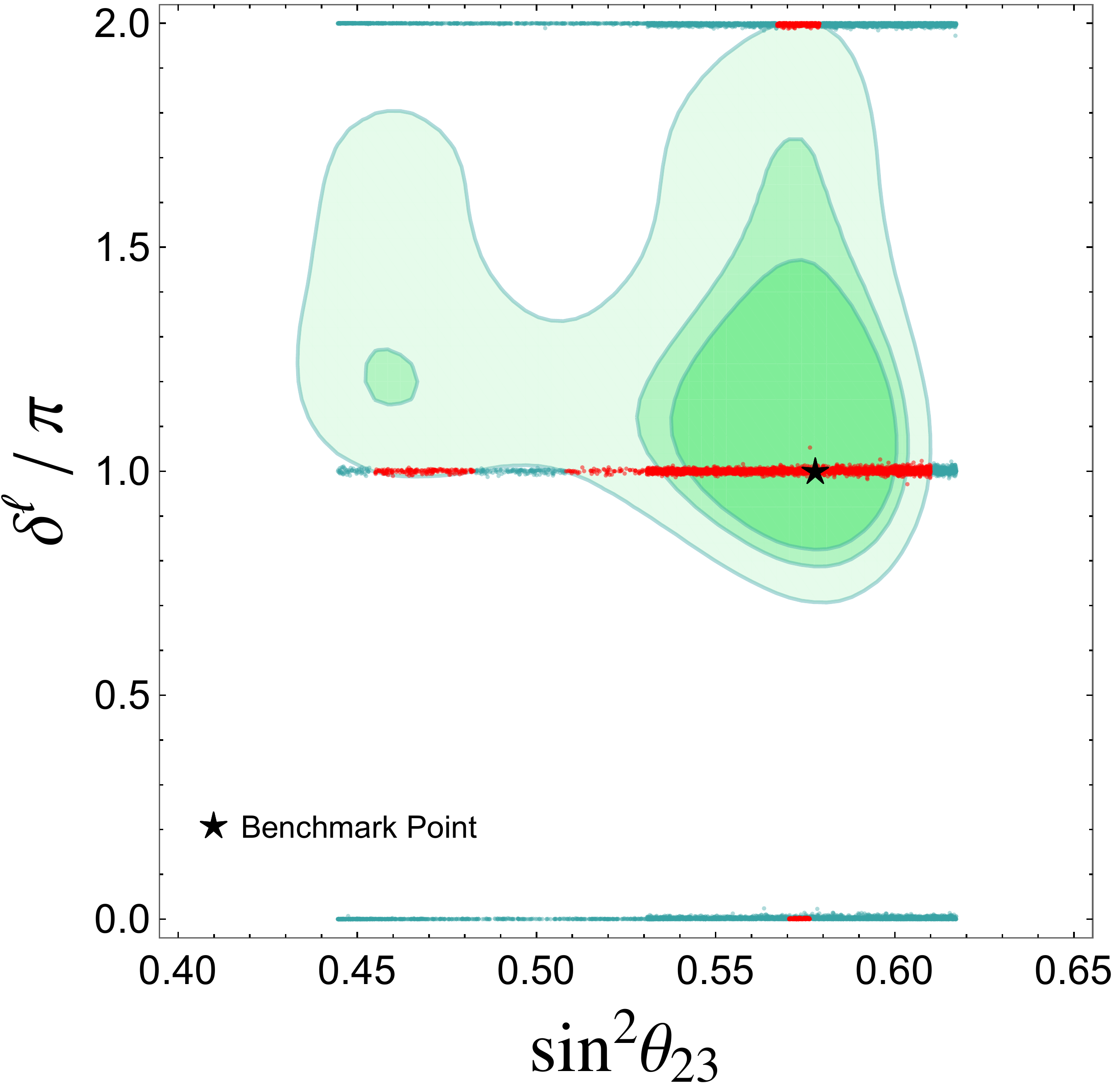}
}}%
    \qquad
    \subfloat[
    Strong correlation between the mass of the lightest neutrino $m^{\nu}_1$ and $\sin ^{2}\theta_{21}$ in this model. This correlation constrains the $m^{\nu}_1$ attainable range.]{{
     \includegraphics[width=2.70in]{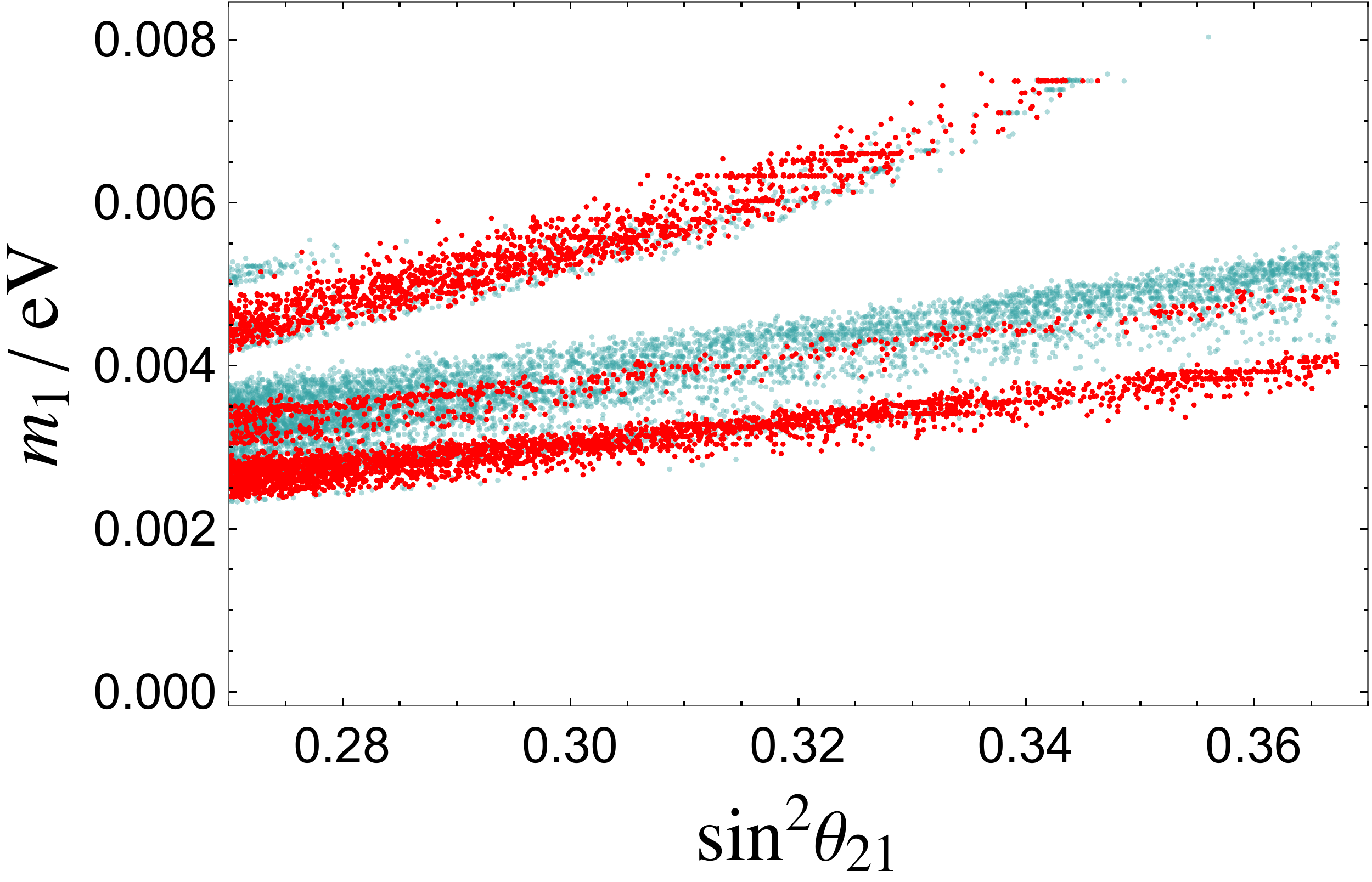}
}}%
    \caption{
  Scatter plots of viable points from our parameter scan.  The red points are inside the allowed region for $\delta^{\ell}$. The coloring of the points matches in both figures. These figures were taken directly from \cite{Bonilla:2023pna}.}%
    \label{fig:CorrelationPlots}%
\end{figure}

In this model, the Yukawa couplings  in Eqs. (\ref{eq:MDirA}), and (\ref{eq:YukEtaCouplings}) can be set to be real without a loss of generality due to the $A_4$ symmetry
\begin{equation}
\{ y_{e},\hspace{2pt}y_{\mu},\hspace{2pt}y_{\tau},\hspace{2pt}y^{\nu}_{1},\hspace{2pt}y^{\nu}_{2},\hspace{2pt}y^{\nu}_{3} \},
    \label{eq:YukawaCouplings}
\end{equation}
Therefore, the discrete dark matter model has two properties concerning CP-symmetry;
the first one is that lepton CP is strictly preserved at the tree-level, thus it can only emerge from the scalar potential's CP-breaking couplings through radiative corrections,
the second one is that from our analysis of the parameter space, we found that assuming CP conservation in the scalar potential is not a viable scenario, 
since in this case the mixing angle between the dark scalars $\eta_2$, $\eta_3$ is fixed (to $\pi/4$) and the model is unable to reproduce the observed oscillation parameters. Hence this model requires CP violation in the scalar sector to be viable. Nonetheless, the results from scanning the parameter space
give values of the lepton CP-phase $\delta^{\ell}$ close to lepton CP-conservation
\begin{equation}
\delta^{\ell} \sim 0,\hspace{2pt} 2 \pi, \qquad \delta^{\ell} \sim \pi\,,
    \label{eq:CPPreserv}
\end{equation}
as shown in Figure \ref{fig:CorrelationPlots} (a). Furthermore, we find a correlation between the lightest neutrino mass $m^{\text{lightest}}_{\nu} = m^{\nu}_1$ and the solar mixing angle ($\sin^2 \theta_21$), and it constrains the value of $m^{\text{lightest}}_{\nu}$ to the range
\begin{equation}
2 \text{ meV} \hspace{2pt} \leq m^{\text{lightest}}_{\nu} \leq  \hspace{2pt} 8 \text{ meV},
\label{eq:LightestMassRange}
\end{equation}
as can be seen in Figure  \ref{fig:CorrelationPlots} (b).
\begin{table}[ht]
	\centering
	\scriptsize
	\begin{tabular}[t]{ |l |c|c c |c| }
		\hline
		\multirow{2}{*}{Observable}& \multicolumn{2}{c}{Data} & & \multirow{2}{*}{Best fit}  \\
		\cline{2-4}
		& Central value & 1$\sigma$ range  &   & \\
		\hline
		$\sin^2\theta_{12}/10^{-1}$ & 3.18 & 3.02 $\to$ 3.34 && $3.03$  \\ 
		$\sin^2\theta_{13}/10^{-2}$ (NO) & 2.200 & 2.138 $\to$ 2.269 && $2.201$  \\  
		$\sin^2\theta_{23}/10^{-1}$ (NO) & 5.74 & 5.60 $\to$ 5.88 && $5.78$  \\
		$\delta^\ell$ $/\pi$ (NO) & 1.08 & 0.96 $\to$ 1.21 && $1.0$  \\
		$\Delta m_{21}^2 / (10^{-5} \, \mathrm{eV}^2 ) $  & 7.50  & 7.30 $\to$ 7.72 && $7.53$  \\
		$\Delta m_{31}^2 / (10^{-3} \, \mathrm{eV}^2) $ (NO) & 2.55  & 2.52 $\to$ 2.57 &&  $2.55$ \\
		$m^{\nu}_{\text{lightest}}$ $/\mathrm{meV}$  (NO) & & & & $ 5.57 $ \\ 
		$m^{\nu}_2$ $/\mathrm{meV}$  & && & $ 10.37$ \\ 
		$m^{\nu}_3$ $/\mathrm{meV}$  & && & $50.87 $ \\
		$ \phi_{12}/ \pi $ & & && $1.5$  \\
		$ \phi_{13}/ \pi$ & & && $1.5$  \\    
		$ \phi_{23}/ \pi$ & & && $1.0$  \\
		$ \langle m_{\beta \beta} \rangle/ \mathrm{eV}$ & & && $3.15\times 10^{-4}$  \\
		$m_e$ $/ \mathrm{MeV}$ & 0.486 &  0.486 $\to$ 0.486 && $0.486$ \\ 
		$m_\mu$ $/  \mathrm{GeV}$ & 0.102 & 0.102  $\to$ 0.102  &&  $0.102$ \\ 
		$m_\tau$ $/ \mathrm{GeV}$ &1.746 & 1.743 $\to$1.747 && $1.746$ \\     	
		\hline
		$v_{\eta_1}/\mathrm{GeV}$ && & &$84.6$ \\
		$v_{H}/\mathrm{GeV}$ &&&& $231.0$\\
         $v/\mathrm{GeV}$ &246 & 246 $\to$ 246 &&  $246$\\
		$M_H/\mathrm{GeV}$  (Higgs)&125.25 & 125.08 $\to$ 125.42 &&  $125.25$ \\	
		\hline
	\end{tabular}
        \caption{
    Model parameters and corresponding observables for this benchmark point.} 
	\label{tab:fit}
\end{table}

\subsection{Dark matter and scalar spectrum}

In this model, the lightest $\mathbb{Z}_2$-odd particle is a dark matter candidate. From the parameter space scan, we found that for viable lepton mixing in this model, the heavy fermion mass $m_N$ is much larger than that of the neutral dark scalars. Hence this model has a scalar as WIMP dark matter candidate with a mass of the order of the electroweak scale. Furthermore, the scalar spectrum is very rich in this model. Since there are four scalar $SU(2)$ doublets in the model, to illustrate it, we summarize the results for the scalar field spectrum in Table  \ref{tab:fitScalars}. More details and results about the analysis of the scalar sector are in \cite{Bonilla:2023pna}.

\begin{table}[ht]
	\centering
	\footnotesize
		\begin{tabular}[t]{ |l |c|c c |c| }
		\hline
		\multirow{2}{*}{Scalar field}& \multicolumn{2}{c}{Data} & & \multirow{2}{*}{Model best fit}  \\
		\cline{2-4}
		& Central value & 1$\sigma$ range  &   & \\
		\hline
		$M_H/\mathrm{GeV}$  (Higgs boson)&125.25 & 125.08 $\to$ 125.42 &&  $125.25$ \\	
		$M_{DM}/\mathrm{GeV}$  (lightest dark scalar)   & & && $60.00$  \\
		$M_{H_{0}}/\mathrm{GeV}$  \text{(Heavy Higgs)}& &  &&  $294.86$ \\	
		$M_{A_0}/\mathrm{GeV}$  \text{(Heavy Pseudoscalar)}   & & && $402.41$  \\
		$M^{+}_{H_0}/\mathrm{GeV}$ \text{(Charged Active)} & & && $313.81$ \\	
		$M_{\chi^+_1}/\mathrm{GeV}$ \text{(Charged Dark)} & & && $482.3$ \\	
		$M_{\chi^+_2}/\mathrm{GeV}$ \text{(Charged Dark)} & & && $341.11$ \\
		$M^{0}_{\chi_1}/\mathrm{GeV}$ \text{(Neutral Dark)} & & && $324.04$ \\
		$M^{0}_{\chi_2}/\mathrm{GeV}$ \text{(Neutral Dark)} & & && $322.44$ \\
		$M^{0}_{\chi_3}/\mathrm{GeV}$ \text{(Neutral Dark)} & & && $82.4$ \\
       		\hline
	\end{tabular}
        \caption{
    Benchmark point values for the masses of the scalar fields in the model.} 
	\label{tab:fitScalars}
\end{table}

\section{Conclusions}
\label{sec:Conclusions}
In this work \cite{Bonilla:2023pna}, we have studied the discrete dark matter model (DDM), which is based on a $A_4$ flavor symmetry in the lepton sector, where the interplay between the seesaw and scotogenic mechanisms arises directly from a remnant symmetry of the spontaneous $A_4$ breaking. This residual symmetry stabilizes a scalar dark matter candidate. 

The model has the unique property among the scoto-seesaw models, the Yukawa couplings ruling both mechanisms are the same. Therefore  tree-level seesaw dominates over the one-loop scotogenic mechanism due to the $A_4$ flavor symmetry. Consequently the model predicts normal ordering (NO) for neutrino masses, favored by neutrino oscillation experiments.  The only source of leptonic CP-violation originates in the scalar potential. Furthermore, the latter is necessary to obtain lepton mixing parameters consistent with the current experimental values. The values obtained for the Dirac CP-phase lie close to the CP-preserving values, as displayed in Figure \ref{fig:CorrelationPlots}.
There is a strong correlation between the mass of the lightest neutrino mass (NO) and the solar mixing angle, which constrains $m^{\nu}_1$ to be between $2\hspace{2pt} \text{meV}\hspace{2pt}\lesssim \hspace{2pt} m^{\nu}_{\text{lightest}} \hspace{2pt} \lesssim \hspace{2pt} 8 \hspace{2pt} \text{meV}$.
The model contains a rich scalar sector with charged and neutral states, accessible to collider experiments. Previous works on the DDM model explore this sector \cite{Hirsch:2010ru,Boucenna:2011tj}.

\begin{acknowledgments} 
Work supported by  the Spanish grants PID2020-113775GB-I00 (AEI/10.13039/501100011033) and Prometeo CIPROM/2021/054 (Generalitat Valenciana), and the Mexican grants CONACYT CB-2017-2018/A1-S-13051 and DGAPA-PAPIIT IN107621 
CB is supported by ANID under the FONDECYT grant “Nu Physics” No. 11201240.
OM is supported by  Programa Santiago Grisolía (No. GRISOLIA/2020/025).
EP is grateful to funding from `C\'atedras Marcos Moshinsky' (Fundaci\'on Marcos Moshinsky).

\end{acknowledgments}

%


\begin{thebibliography}{99}
\small
\providecommand{\url}[1]{\texttt{#1}}
\providecommand{\urlprefix}{URL }
\providecommand{\eprint}[2][]{\url{#2}}


\bibitem{Bonilla:2023pna}
C. Bonilla, J. Herms, O. Medina and E. Peinado,
 \emph{{Discrete dark matter mechanism as the source of
neutrino mass scales}},
 \MYhref[eprintLinks]{https://arxiv.org/abs/2301.10811}{{\ttfamily
  arXiv:2301.10811 [hep-ph]}}.
  
  \bibitem{deSalas:2020pgw}
P.~F. de~Salas et~al., \emph{{2020 global reassessment of the neutrino
  oscillation picture}},
  \MYhref[journalLinks]{http://dx.doi.org/10.1007/JHEP02(2021)071}{JHEP
  }\MYhref[journalLinks]{http://dx.doi.org/10.1007/JHEP02(2021)071}{\textbf{02}
  (2021) 071}, \MYhref[eprintLinks]{http://arxiv.org/abs/2006.11237}{{\ttfamily
  arXiv:2006.11237 [hep-ph]}}.
    
   \bibitem{Hirsch:2010ru}
M.~Hirsch, S.~Morisi, E.~Peinado and J.~W.~F. Valle, \emph{{Discrete dark
  matter}},
  \MYhref[journalLinks]{http://dx.doi.org/10.1103/PhysRevD.82.116003}{Phys.
  Rev. D
  }\MYhref[journalLinks]{http://dx.doi.org/10.1103/PhysRevD.82.116003}{\textbf{82}
  (2010) 116003},
  \MYhref[eprintLinks]{http://arxiv.org/abs/1007.0871}{{\ttfamily
  arXiv:1007.0871 [hep-ph]}}.
  
  \bibitem{Boucenna:2011tj}
M.~S. Boucenna et~al., \emph{{Phenomenology of Dark Matter from $A_4$ Flavor
  Symmetry}},
  \MYhref[journalLinks]{http://dx.doi.org/10.1007/JHEP05(2011)037}{JHEP
  }\MYhref[journalLinks]{http://dx.doi.org/10.1007/JHEP05(2011)037}{\textbf{05}
  (2011) 037}, \MYhref[eprintLinks]{http://arxiv.org/abs/1101.2874}{{\ttfamily
  arXiv:1101.2874 [hep-ph]}}.

\bibitem{Schechter:1980gr}
J.~Schechter and J.~W.~F. Valle, \emph{{Neutrino Masses in SU(2) x U(1)
  Theories}},
  \MYhref[journalLinks]{http://dx.doi.org/10.1103/PhysRevD.22.2227}{Phys. Rev.
  D
  }\MYhref[journalLinks]{http://dx.doi.org/10.1103/PhysRevD.22.2227}{\textbf{22}
  (1980) 2227}.
 
  \bibitem{Ma:2006km}
E.~Ma, \emph{{Verifiable radiative seesaw mechanism of neutrino mass and dark
  matter}},
  \MYhref[journalLinks]{http://dx.doi.org/10.1103/PhysRevD.73.077301}{Phys.
  Rev. D
  }\MYhref[journalLinks]{http://dx.doi.org/10.1103/PhysRevD.73.077301}{\textbf{73}
  (2006) 077301},
  \MYhref[eprintLinks]{http://arxiv.org/abs/hep-ph/0601225}{{\ttfamily
  arXiv:hep-ph/0601225}}.
  
\bibitem{Escribano:2020iqq}
P.~Escribano, M.~Reig and A.~Vicente, \emph{{Generalizing the Scotogenic
  model}},
  \MYhref[journalLinks]{http://dx.doi.org/10.1007/JHEP07(2020)097}{JHEP
  }\MYhref[journalLinks]{http://dx.doi.org/10.1007/JHEP07(2020)097}{\textbf{07}
  (2020) 097}, \MYhref[eprintLinks]{http://arxiv.org/abs/2004.05172}{{\ttfamily
  arXiv:2004.05172 [hep-ph]}}.

\bibitem{Rojas:2018wym}
N.~Rojas, R.~Srivastava and J.~W.~F. Valle, \emph{{Simplest Scoto-Seesaw
  Mechanism}},
  \MYhref[journalLinks]{http://dx.doi.org/10.1016/j.physletb.2018.12.014}{Phys.
  Lett. B
  }\MYhref[journalLinks]{http://dx.doi.org/10.1016/j.physletb.2018.12.014}{\textbf{789}
  (2019) 132--136},
  \MYhref[eprintLinks]{http://arxiv.org/abs/1807.11447}{{\ttfamily
  arXiv:1807.11447 [hep-ph]}}.

\bibitem{Aranda:2018lif}
A.~Aranda, C.~Bonilla and E.~Peinado, \emph{{Dynamical generation of neutrino
  mass scales}},
  \MYhref[journalLinks]{http://dx.doi.org/10.1016/j.physletb.2019.01.068}{Phys.
  Lett. B
  }\MYhref[journalLinks]{http://dx.doi.org/10.1016/j.physletb.2019.01.068}{\textbf{792}
  (2019) 40--42},
  \MYhref[eprintLinks]{http://arxiv.org/abs/1808.07727}{{\ttfamily
  arXiv:1808.07727 [hep-ph]}}.

\bibitem{Ibarra:2011gn}
A.~Ibarra and C.~Simonetto, \emph{{Understanding neutrino properties from
  decoupling right-handed neutrinos and extra Higgs doublets}},
  \MYhref[journalLinks]{http://dx.doi.org/10.1007/JHEP11(2011)022}{JHEP
  }\MYhref[journalLinks]{http://dx.doi.org/10.1007/JHEP11(2011)022}{\textbf{11}
  (2011) 022}, \MYhref[eprintLinks]{http://arxiv.org/abs/1107.2386}{{\ttfamily
  arXiv:1107.2386 [hep-ph]}}.

\bibitem{Barreiros:2020gxu}
D.~M. Barreiros, F.~R. Joaquim, R.~Srivastava and J.~W.~F. Valle,
  \emph{{Minimal scoto-seesaw mechanism with spontaneous CP violation}},
  \MYhref[journalLinks]{http://dx.doi.org/10.1007/JHEP04(2021)249}{JHEP
  }\MYhref[journalLinks]{http://dx.doi.org/10.1007/JHEP04(2021)249}{\textbf{04}
  (2021) 249}, \MYhref[eprintLinks]{http://arxiv.org/abs/2012.05189}{{\ttfamily
  arXiv:2012.05189 [hep-ph]}}.
  
\bibitem{Barreiros:2022aqu}
D.~M. Barreiros, H.~B. Camara and F.~R. Joaquim, \emph{{Flavour and dark matter
  in a scoto/type-II seesaw model}},
  \MYhref[journalLinks]{http://dx.doi.org/10.1007/JHEP08(2022)030}{JHEP
  }\MYhref[journalLinks]{http://dx.doi.org/10.1007/JHEP08(2022)030}{\textbf{08}
  (2022) 030}, \MYhref[eprintLinks]{http://arxiv.org/abs/2204.13605}{{\ttfamily
  arXiv:2204.13605 [hep-ph]}}.

\bibitem{Mandal:2021yph}
S.~Mandal, R.~Srivastava and J.~W.~F. Valle, \emph{{The simplest scoto-seesaw
  model: WIMP dark matter phenomenology and Higgs vacuum stability}},
  \MYhref[journalLinks]{http://dx.doi.org/10.1016/j.physletb.2021.136458}{Phys.
  Lett. B
  }\MYhref[journalLinks]{http://dx.doi.org/10.1016/j.physletb.2021.136458}{\textbf{819}
  (2021) 136458},
  \MYhref[eprintLinks]{http://arxiv.org/abs/2104.13401}{{\ttfamily
  arXiv:2104.13401 [hep-ph]}}.

\bibitem{Ganguly:2022qxj}
J.~Ganguly, J.~Gluza and B.~Karmakar, \emph{{Common origin of
  \ensuremath{\theta}$_{13}$ and dark matter within the flavor symmetric
  scoto-seesaw framework}},
  \MYhref[journalLinks]{http://dx.doi.org/10.1007/JHEP11(2022)074}{JHEP
  }\MYhref[journalLinks]{http://dx.doi.org/10.1007/JHEP11(2022)074}{\textbf{11}
  (2022) 074}, \MYhref[eprintLinks]{http://arxiv.org/abs/2209.08610}{{\ttfamily
  arXiv:2209.08610 [hep-ph]}}.


\end{thebibliography}
\end{document}